%% file: HPNP2015_Moretti.tex
\documentclass[a4paper]{revtex4}
\usepackage{graphicx}
\usepackage{fancyhdr}
\usepackage{amsmath}
\usepackage{color}

\begin{document}

\input macros_gen

%Title of paper
\title{A light NMSSM pseudoscalar Higgs boson at the LHC Run 2}

% Repeat the \author .. \affiliation  etc. as needed
%
% \affiliation command applies to all authors since the last
% \affiliation command. The \affiliation command should follow the
% other information

\author{N-E. Bomark$^a$}
\author{S. Moretti$^{b,*}$}
\author{S. Munir$^{c,d}$}
\author{L. Roszkowski$^{a,e}$}

\affiliation{$^a$National Centre for Nuclear Research,  Ho{\. z}a 69, 00-681 Warsaw, Poland}
\affiliation{$^b$School of Physics \& Astronomy, University of Southampton, Southampton SO17 1BJ, UK}
\affiliation{$^c$Asia Pacific Center for Theoretical Physics, San 31, Hyoja-dong, Nam-gu, Pohang 790-784, Republic of Korea}
\affiliation{$^d$Department of Physics and Astronomy, Uppsala University, Box 516, SE-751 20 Uppsala, Sweden}
\affiliation{$^e$Department of Physics \& Astronomy, University of Sheffield, Western Bank, Sheffield S10 2TN, UK}

\begin{abstract}
We revisit the light pseudoscalar $A_1$ in the Next-to-Minimal Supersymmetric Standard Model
(NMSSM) with partial universality at some high unification scale in order to delineate the 
parameter space regions consistent with  up-to-date theoretical
and experimental constraints and examine to what extent this state can be
probed by the Large Hadron Collider (LHC) during Run 2. We find that it can be accessible through a variety of signatures  proceeding via  $A_1\to \tau^+\tau^-$ and/or
$b\bar b$, the former assuming hadronic decays and the latter two $b$-tags within  a fat jet or two separate slim ones. Herein,  the light pseudoscalar state is produced
 from a heavy Higgs
boson decay in either pairs  or
singly in association with a $Z$ boson (in turn decaying into electrons/muons).     
\end{abstract}

%\maketitle must follow title, authors, abstract
\maketitle

\thispagestyle{fancy}

% body of paper here - Use proper section commands
% References should be done using the  \cite, \ref, and \label commands
% Put \label in argument of \section for cross-referencing
%\section{\label{}}

%%%%%%%%%%%%%%%%%%%%%%%%%%%%%%%%%%
\section{Introduction}

In this report we analyse in detail some of the  processes
that yield  sizeable event rates and could potentially lead to
the detection of a light $A_1$ at the LHC with $\sqrt{s} =14$ TeV within the NMSSM~\cite{Ellwanger:2009dp}.
We perform parameter scans of this scenario with partial
universality at the Grand Unification Theory (GUT) scale to find regions where a light,
$\lesssim 150$ GeV, $A_1$ can be obtained. In these scans we require
the mass of the SM-like Higgs state discovered at the LHC, henceforth denoted by $H_{\rm SM}$, to lie around 125 GeV and its signal
rates in the $\gamma\gamma$ and $ZZ$ channels to be
consistent with the SM expectations. We study in detail the
two possibilities, ${H_{\rm SM}} = H_1$ and
 ${H_{\rm SM}} = H_2$, as two separate cases. 
(Recall that the neutral Higgs spectrum of the NMSSM includes three CP-even states, $H_{1,2,3}$, and two CP-odd ones, $A_{1,2}$, wherein an
increasing numerical label represents an heavier state.)
Moreover, we assume the $A_1$ to be produced via the
decay of a heavy scalar Higgs boson of the model, the latter induced by $gg$ fusion.
(As finally established in~\cite{Bomark:2014gya},
although some scope was demonstrated for single $A_1$ production in association with a $b\bar{b}$ prior to LHC Higgs data~\cite{Almarashi:2012ri}, this channel no longer carries any promise. Also, note that the scope of Vector Boson Fusion (VBF) and Higgs-strahlung is also currently being re-assessed in the light of the same experimental results~\cite{preparation}.)
In particular, we include the two 
intermediate channels  $A_1A_1$ and $A_1Z$ while the decaying heavier Higgs boson can
be any of the three neutral scalars. %, $H_1$, $H_2$ and $H_3$. 
$A_1$'s thus produced decay into either $b\bar b$ or (fully hadronic)
$\tau^+\tau^-$ pairs. The former decay  is always the
dominant one as the ratio of the branching ratios (BRs) for these modes
is given approximately by the ratio of the $b$ and the $\tau$ masses squared,
but the latter decay can be equally important due to a relatively smaller $\tau^+\tau^-$ background.
In case of the $A_1Z$ decay channel, we only consider the leptonic
($e^+e^-$ and $\mu^+\mu^-$) decays of the $Z$ boson.

To study the prospects for the discovery of an $A_1$ at the LHC in all
these production and decay channels, we employ
hadron level Monte Carlo (MC) simulations. We perform a detailed
signal-to-background analysis for each process of interest, employing
jet substructure methods for detecting the
$b$ quarks originating from an $A_1$ decay, assuming two $b$-tags in either two single $b$-jets or one fat $b$-jet.
In particular, notice that, in case of a decaying SM-like Higgs state, the mass measurement of 
$\sim 125$\gev\ serves as an 
important kinematical handle for all signatures. Removing this condition reduces the sensitivity by a factor of $2$ to $3$. To recap,
the $A_1A_1$ pair thus produced 
decays into the $b\bar b b\bar b$ (4$b$), $b\bar b\tau^+\tau^-$ (2$b$2$\tau$) and
$\tau^+\tau^- \tau^+\tau^-$ (4$\tau$) final state combinations while
 in the case of $A_1Z$ production we will be looking at 
 $b\bar b\ell^+\ell^-$ ($2b2\ell$) and $\tau^+\tau^-\ell^+\ell^-$ ($2\tau2\ell$), wherein $\ell=e,\mu$.

\section{\label{model} Model setup and scans}

In order to remedy the proliferation of parameters typical of realistic models of Supersymmetry (SUSY), one
usually invokes some kind of unification of these at high energy scales, typical of GUTs.
However,  as noted in~\cite{Kowalska:2012gs}, the fully constrained
NMSSM (where all scalar and fermion masses as well as dimensionful
couplings are unified, respectively, into three separate parameters) 
struggles to achieve the correct mass for the assumed SM-like Higgs
boson, particularly in the presence of the latest theoretical and  experimental
constraints.  In order to bypass this,
the strict unification conditions mentioned above need to be relaxed.
In a partially unconstrained version of the NMSSM
the soft masses of the Higgs fields, $m_{H_u}$, $m_{H_d}$ and $m_S$,
are taken as independent (from $m_0$)
parameters at the GUT scale. Through the minimisation
conditions of the Higgs potential these three soft masses can then be
traded at the Electro-Weak (EW) scale for the parameters \kap, \mueff\ and
\tanb. Similarly, the soft trilinear coupling
parameters \alam\ and \akap, though still input at the GUT
scale, are also taken as independent (from \azero). The model is thus defined in terms of the following nine continuous input parameters:
\begin{center}
\mzero, \mhalf, \azero, \tanb, \lam, \kap, \mueff, \alam, \akap,
\end{center}
where $\tanb \equiv v_u/v_d$, with $v_u$ being the Vacuum Expectation Value
(VEV) of the $u$-type
Higgs doublet and $v_d$ that of the $d$-type one. This version of the
model serves as a good approximation of the most general
EW-scale NMSSM as far as the Higgs sector dynamics is concerned.
We, therefore, adopt it here  to
analyse the phenomenology of the light pseudoscalar and we refer to
it as the CNMSSM-NUHM, where NUHM stands for
Non-Universal (soft) Higgs Masses.

We scanned the CNMSSM-NUHM parameter space given in Tab.\,\ref{tab:params} to search for
points  giving $m_{A_1} \lesssim 150$\gev. We used the publicly available package 
NMSSMTools-v4.2.1~\cite{NMSSMTools} for computation of the 
mass, coupling and BR spectrum of the Higgs bosons for each
model point.  In our scans we imposed the following  constraints from 
$b$-physics, based on~\cite{Beringer:1900zz}, and dark matter relic density
measurements, from~\cite{Ade:2013zuv}, as
\begin{itemize}
\item $\brbsmumu = (3.2~(\pm 10\%~{\rm theoetical~error})\pm1.35) \times 10^{-9}$,
\item $\brbutaunu = (1.66\pm 0.66 \pm 0.38) \times 10^{-4}$,
\item $\brbxsgamma = (3.43\pm 0.22 \pm0.21) \times 10^{-4}$,
\item $\Omega_\chi h^2 < 0.131~(0.119+10\%~{\rm theoretical~error})$.
\end{itemize}
Exclusion limits from the LEP and LHC Higgs boson searches were also tested
against using the HiggsBounds-v4.1.3~\cite{Bechtle:2013wla} package.
Finally, from NMSSMTools we obtained the signal rates of the $H_{\rm SM}$ state, defined
for a given decay channel $X$ as
\begin{equation}
\label{eq:rggh}
R_X \equiv  \frac{\sigma(gg\rightarrow H_2)\times {\rm BR}(H_i\rightarrow
  X)}{\sigma(gg\rightarrow h_{\rm SM})\times {\rm BR}(h_{\rm SM} \rightarrow X)}\,,
\end{equation}
where ${h_{\rm SM}}$ is the true SM Higgs boson, which we required (for $X=\gamma\gamma,\,ZZ$) to lie within 
the measured $\pm 1\sigma$ ranges of the corresponding experimental
quantities%$\mu_X$ 
\begin{equation}
\label{eq:CMS-mu}\mu_{\gamma \gamma} = 1.13 \pm 0.24\,,~~~\mu_{ZZ} =
1.0 \pm 0.29\,~~{\rm and}~~
\end{equation}
\begin{equation}
\label{eq:ATLAS-mu}
\mu_{\gamma \gamma} = 1.57^{+0.33}_{-0.28}\,, ~~~\mu_{ZZ} = 1.44^{+0.40}_{-0.35}\,,
\end{equation}
 provided by the CMS~\cite{CMS-PAS-HIG-14-009} and
 ATLAS~\cite{ATLAS-CONF-2014-009} collaborations, respectively.
 The red and blue points in the forthcoming figures are the ones for which the calculated $R_X$
lies within the range of $\mu_X$ measured by CMS
and ATLAS, respectively, while the
green points are the `unfiltered' ones for which neither of these two
constraints are satisfied.

\begin{table}[tb]
\begin{center}
\caption{The CNMSSM-NUHM input parameters and their scanned ranges.}
\label{tab:params}
\begin{tabular}{|c|c|c|c|c|}
\hline
 Parameter & \mzero\,(GeV)  & \mhalf\,(GeV) & \azero\,(GeV) & \mueff\,(GeV)  \\
\hline  
Range  & 200 -- 2000  & 100 -- 1000 & $-3000$ -- 0 & 100 -- 200 \\
\hline
\hline
\tanb & \lam  & \kap  & \alam\,(GeV)  & \akap\,(GeV) \\
\hline 
1 -- 6 & 0.4 -- 0.7  & 0.01 -- 0.7 & $-500$ -- 500  & $-500$ -- 500 \\
\hline
\end{tabular}
\end{center}
\end{table}

\section{\label{analysis} Signal-to-background Analysis}

Following the scans, we carried out a dedicated
signal-to-background analysis based on MC event
generation for $pp$ collisions at 14\,TeV and variable integrated luminosity, for each process of interest. 
Using the program SuSHi-v1.1.1~\cite{Harlander:2012pb}, we
first calculated the $gg$ fusion production cross section
of a SM Higgs boson with the same mass as as that of our $H_i$. 
This cross section was then rescaled using the $ggH_i$
reduced coupling in the NMSSM and multiplied by the relevant BRs of
the $H_i$'s, all of which are obtained from NMSSMTools. 
The backgrounds, which include  $pp\to 4b$,
$pp\to 2b2\tau$, $pp\to 4\tau$,
$pp\to Z 2b$ and $pp\to Z2\tau$, were computed
with MadGraph5\_aMC$@$NLO~\cite{Alwall:2014hca}. Both  signal and 
background for each process were hadronised and
fragmented using Pythia 8.180~\cite{Sjostrand:2007gs} interfaced with
 FastJet-v3.0.6~\cite{Cacciari:2011ma} for jet clustering. The parton-level acceptance cuts
used in the event generation in MadGraph are:
(i) $|\eta|<$ 2.5 and $p_T>15$\gev\ for all final state objects;
(ii) $\Delta R \equiv \sqrt{(\Delta\eta)^2+(\Delta\phi)^2}>0.2$ for all $b$-quark pairs;
(iii) $\Delta R >0.4$ for all other pairs of final state objects
(where $p_T$, $\eta$, $\phi$ are the transverse momentum,
pseudorapidity and azimuthal angle, respectively). 

As intimated, our use of the jet substructure method~\cite{Butterworth:2008iy}
 implies that we have three possible signatures for a decaying $A_1$: one
 fat jet, two single $b$-jets and two $\tau$-jets. The fat jet
 analysis, which assumes boosted $b$-quarks, allows one to obtain 
much higher sensitivities, particularly
 for large masses of the decaying Higgs bosons. Notice that a key ingredient of this selection is the retention of two $b$-tags in both
cases of a single fat $b$-jet and two slim $b$-jets. Failing this, i.e., if only one $b$-jet were to be tagged instead, the list of backgrounds would dramatically increase, 
in the form of QCD processes also including light-quark and gluon jets. 

\begin{figure}[tbp]
\centering
{%
\label{fig:-a}%
\includegraphics*[width=7.3cm]{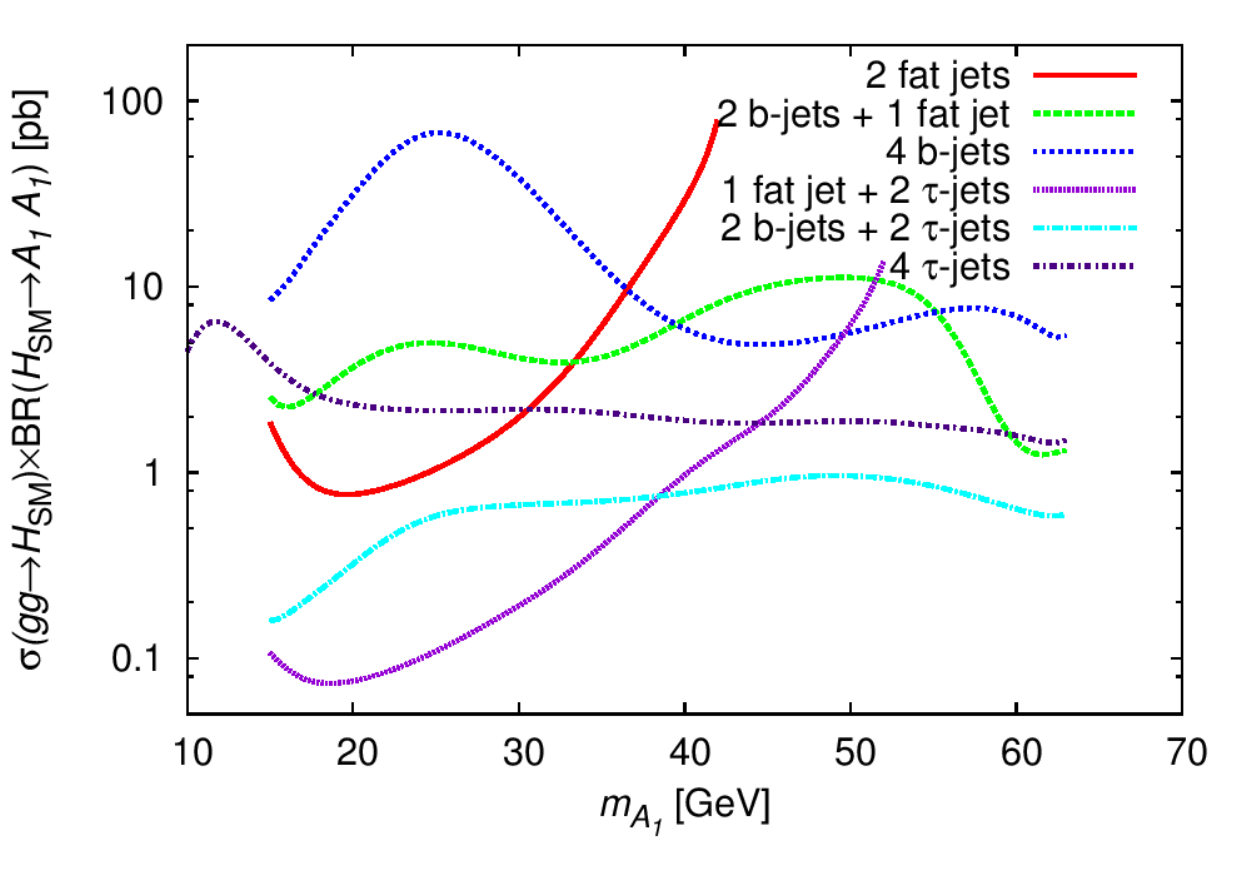}
}%
\hspace{0.5cm}%
{%
\label{fig:-b}%
\includegraphics*[width=7.3cm]{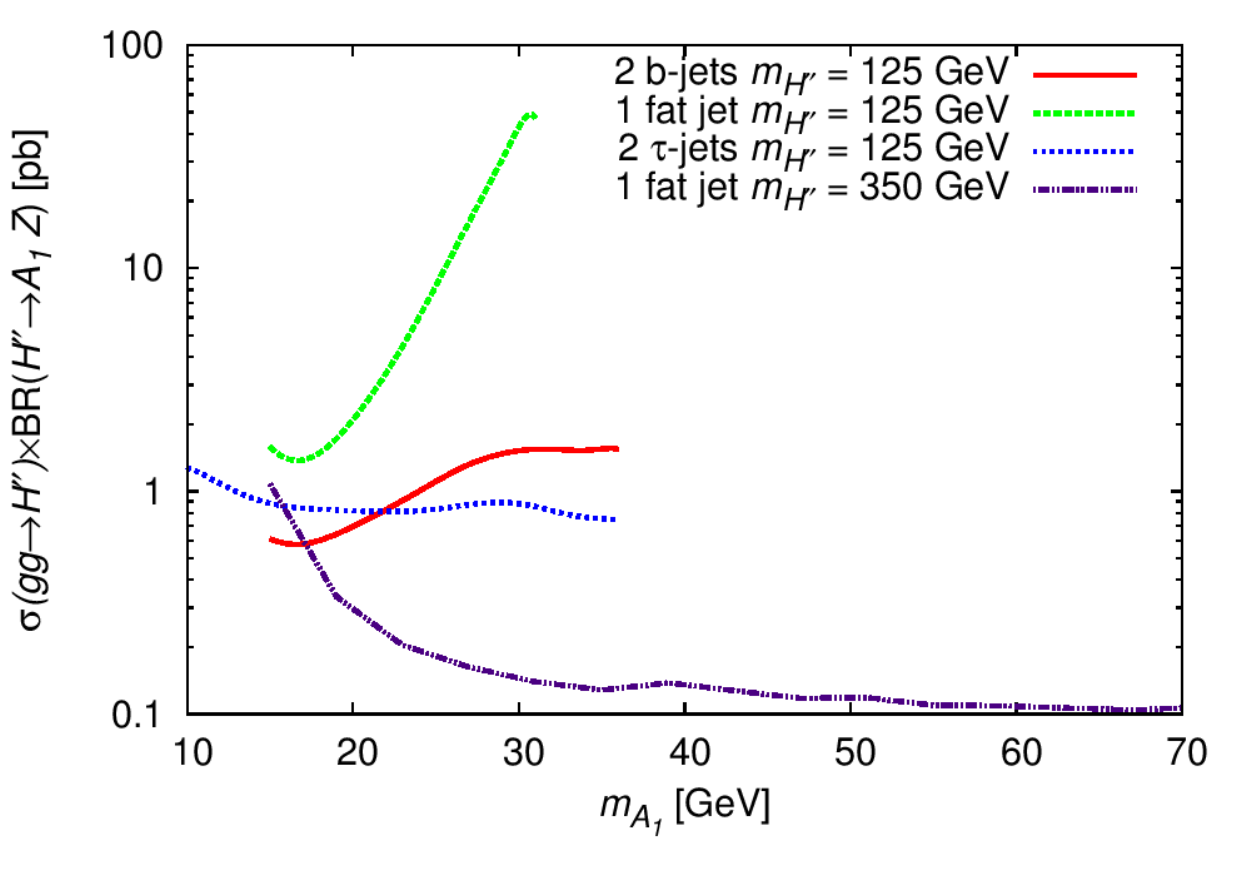}
}%
\caption{Expected experimental sensitivities as functions of $m_{A_1}$
  in various possible final state combinations for (a)
$gg \to H_{\rm SM}\rightarrow A_1A_1$ and (b) $gg \to H_i\rightarrow A_1
Z$.}
\label{fig:Sens}
%\end{center}
\end{figure}

In general, we see in  Fig.~\ref{fig:Sens}(b) that the fat jet analysis
can be very effective when the $A_1$ is much lighter than the $H_i$ ($i=1,2,3$),
but gets worse as $m_{A_1}$ increases and, in fact, soon becomes
relatively useless (the corresponding curves are thus cut off at
the mass above which the analysis becomes ineffective).
This is due to the fact that the fat jet analysis assumes boosted $b$-quark pairs.
One can
also see (especially in the curve with $m_{H'}=350$\gev; hereafter
$H''$ refers to any of the three CP-even Higgs state directly produced in $gg$
fusion while $H'$ refers to the two states other than the
$H_{\rm SM}$ in a given case) that, if
the ${A_1}$ mass becomes too small compared to the $H_i$ mass, the
sensitivity diminishes due to the $b$-jets becoming too collinear
to be separable even with jet substructure methods.
In the upper end, the cut-offs (for sensitivity curves other than those
relying on the fat jet analysis) are determined by the
kinematical upper limit for the given channel,
i.e., $m_{A_1}\approx 62.5$\gev\ for $H_{\rm SM}\to A_1A_1$
and $m_{A_1}\approx 35$\gev\ for $H_{\rm SM}\to A_1Z$.

We finally calculated
the expected cross sections for the signal processes which
yield $S/\sqrt{B}>5$ for three benchmark accumulated luminosities at
the LHC, $\mathcal{L}=30$/fb,\,300/fb and 3000/fb, in various final
state combinations, as functions of $m_{A_1}$. Notice that, in
order to keep the figures readable, in the following section we
will only show the curves corresponding to the analyses with the
highest sensitivities for a given
channel.

\section{\label{results} Results}

\begin{figure}[tbp]
%\begin{center}
%\vspace*{-1cm}
%\begin{tabular}{cc}
\centering
{%
\label{fig:-a}%
%\hspace{-1.cm}%
\includegraphics*[width=7.7cm]{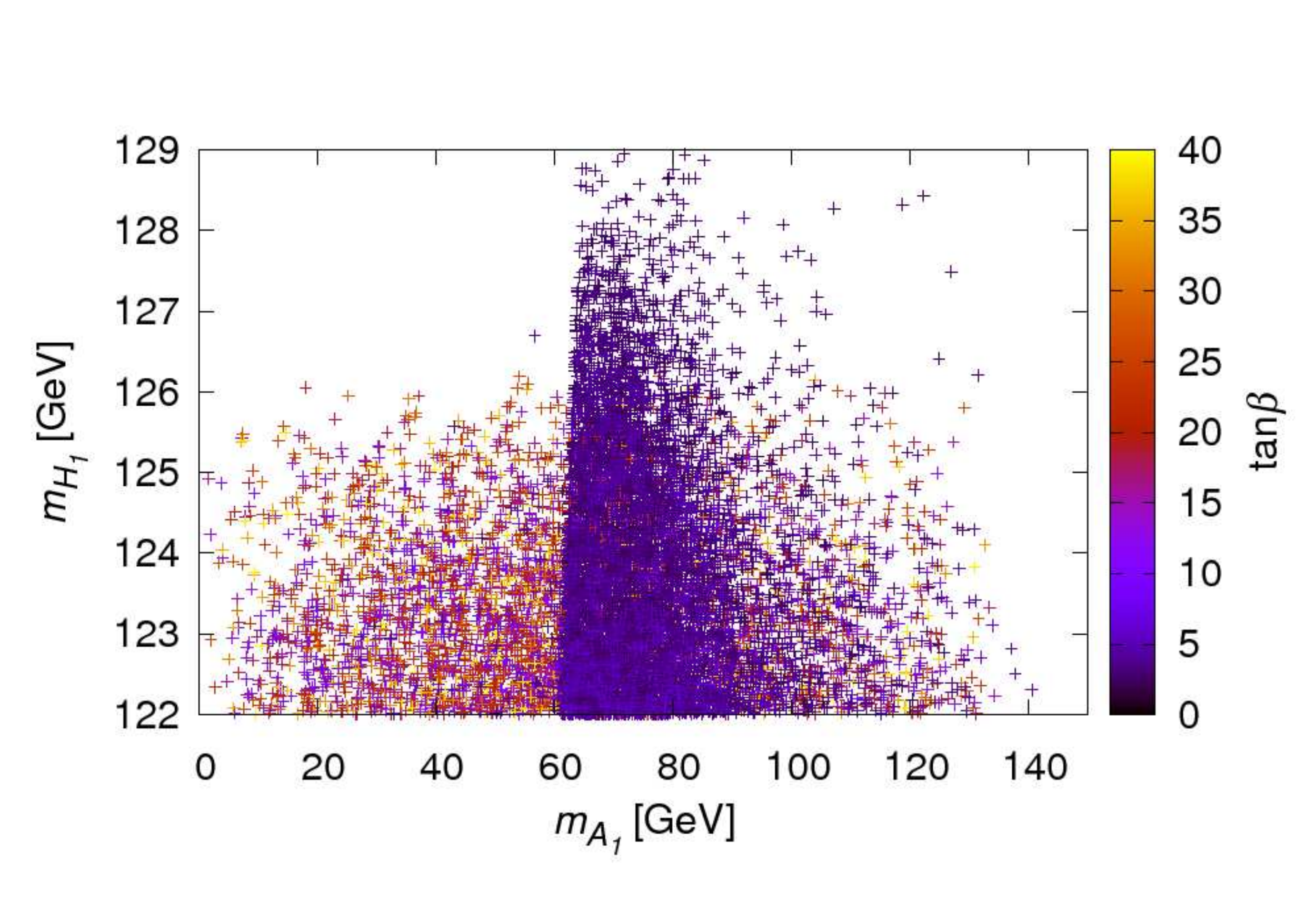}
%\vspace*{-0.5cm}
}%
%\hspace{-0.2cm}%
%\vspace*{-1cm}
{%
\label{fig:-b}%
\includegraphics*[width=7.7cm]{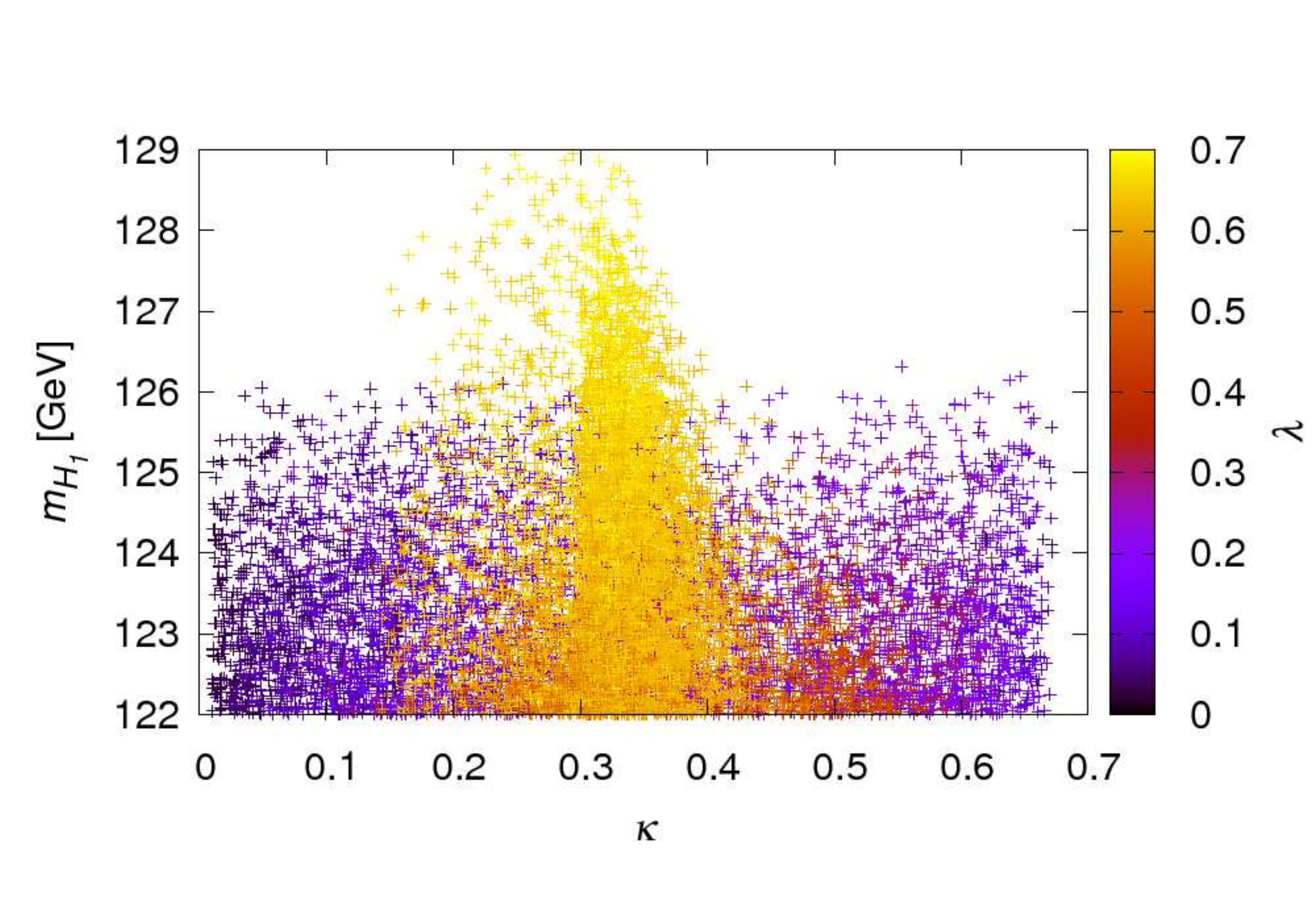}
}%
%(a) & (b) \\
%\end{tabular}
\caption[]{Case with $H_{\rm SM} = H_1$: (a) Mass of $H_1$ vs. that of
  $A_1$, with the heat map
  showing the distribution of \tanb; (b) $m_{H_1}$ as a
  function of the parameter \kap, with the heat map showing the distribution of the coupling \lam.}
\label{fig:h1params1}
%\end{center}
\end{figure}

In the NMSSM, $H_1$ and $H_2$ can  both have masses around $125$\gev\ and
SM-like properties, thereby alternatively playing the role of
$H_{\rm SM}$. A SM-like $H_1$ with mass around 125\gev\ can be obtained over wide
regions of the CNMSSM-NUHM parameter space defined above. However, the
additional requirement of $m_{A_1} \lesssim 150$\gev\ significantly
changes this picture. In Fig.~\ref{fig:h1params1}(a) we show the
distribution of the mass of $H_1$ against that of $A_1$ for the points
obtained in our scans assuming  $H_{\rm SM} = H_1$. We
 allow a rather wide range of $m_{H_{\rm SM}}$, 
$122\gev - 129\gev$, in order to take into account the experimental
as well possibly large theoretical uncertainties in its model
prediction. The heat map in the figure corresponds to the parameter \tanb. One can
see a particularly dense population of points for $\tanb \sim
1-6$ in the figure, with the mass of $H_1$ reaching
comparatively larger values than elsewhere. However, $m_{A_1}$ for
such points almost never falls below $\sim 60$\gev.
In Fig.~\ref{fig:h1params1}(b) we show $m_{H_1}$ as a function of the
coupling \kap, with the heat map corresponding to the coupling
\lam. Again there is a clear strip of points with $\lam \gtrsim
0.6$ (and $\kap \sim 0.15 - 0.5$) for which
$m_{H_1}$ can be as high as 129\gev. The
rest of the points, corresponding to smaller \lam\ and larger \tanb,
can barely yield $m_{H_1}$ in excess of 126\gev. The reason for the behaviour of
$m_{H_1}$ observed in these figures is well explained in Ref.~\cite{Bomark:2014gya} and is essentially
attributable to the expression of  $m_{H_1}$ in terms of the CNMSSM-NUHM input parameters. 

%Despite the above (rather wide) $m_{H_{\rm SM}}$ range was introduced  to take into account the experimental
%as well possibly large theoretical uncertainties in the model
%prediction of $m_{H_{\rm SM}}$, it is clear from these preliminar
%scans that SM-like Higgs mass measurements converging towards the 125 GeV value render
%the SM-like $H_1$ solution of less viability with respect to the $H_2$ one, as made evident by the corresponding Figs. \ref{fig:h2params1}(a) and (b),
%wherein such a central value is more naturally obtained (for $H_2$).
%Also, the same scans make clear that the $H_2$ solution further
%yields small values for $A_1$ more naturally than the $H_1$ case. 

\begin{figure}[tbp]
\centering
{%
\label{fig:-a}%
\includegraphics*[width=7.7cm]{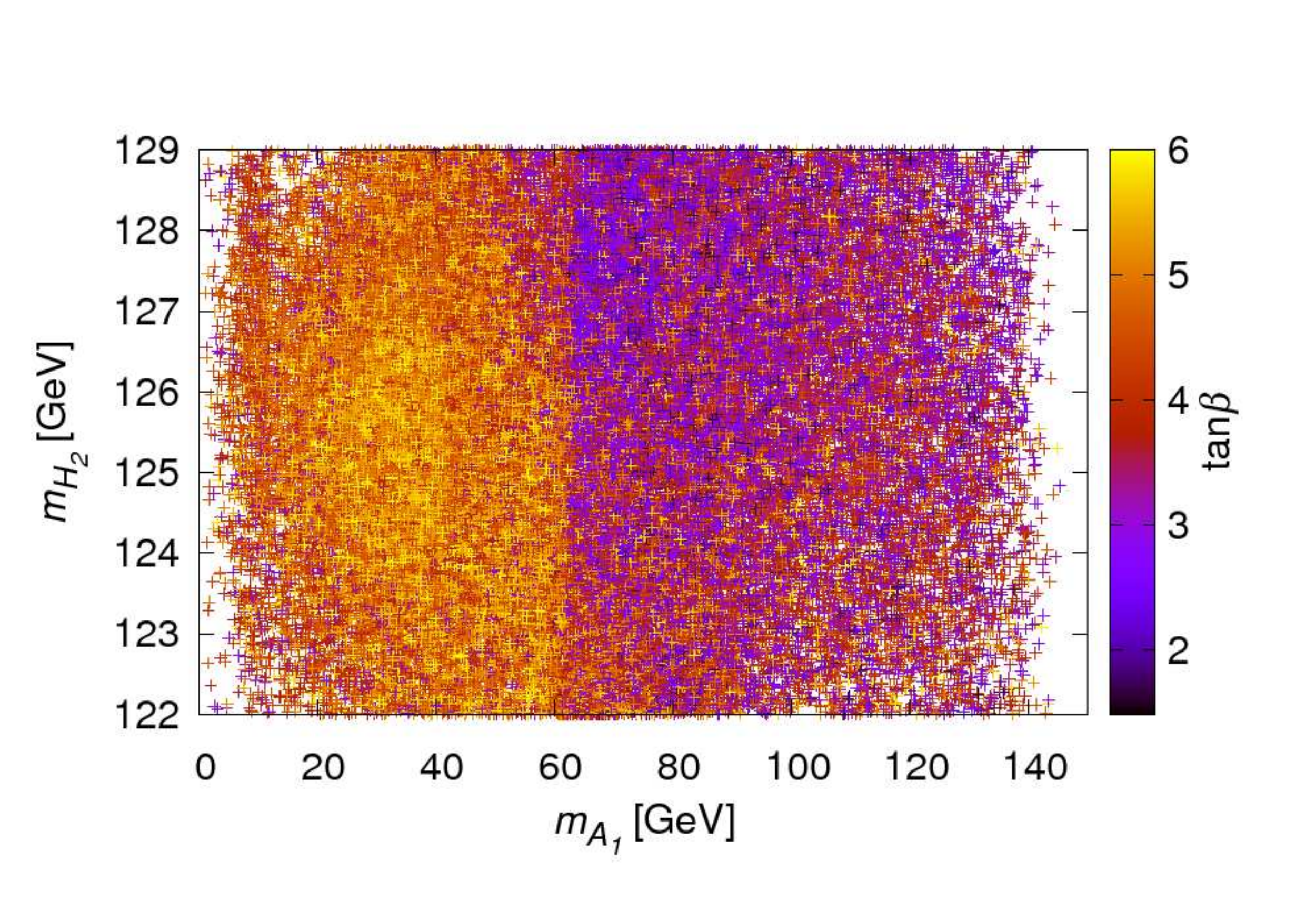}
}%
%\hspace{0.5cm}%
{%
\label{fig:-b}%
\includegraphics*[width=7.7cm]{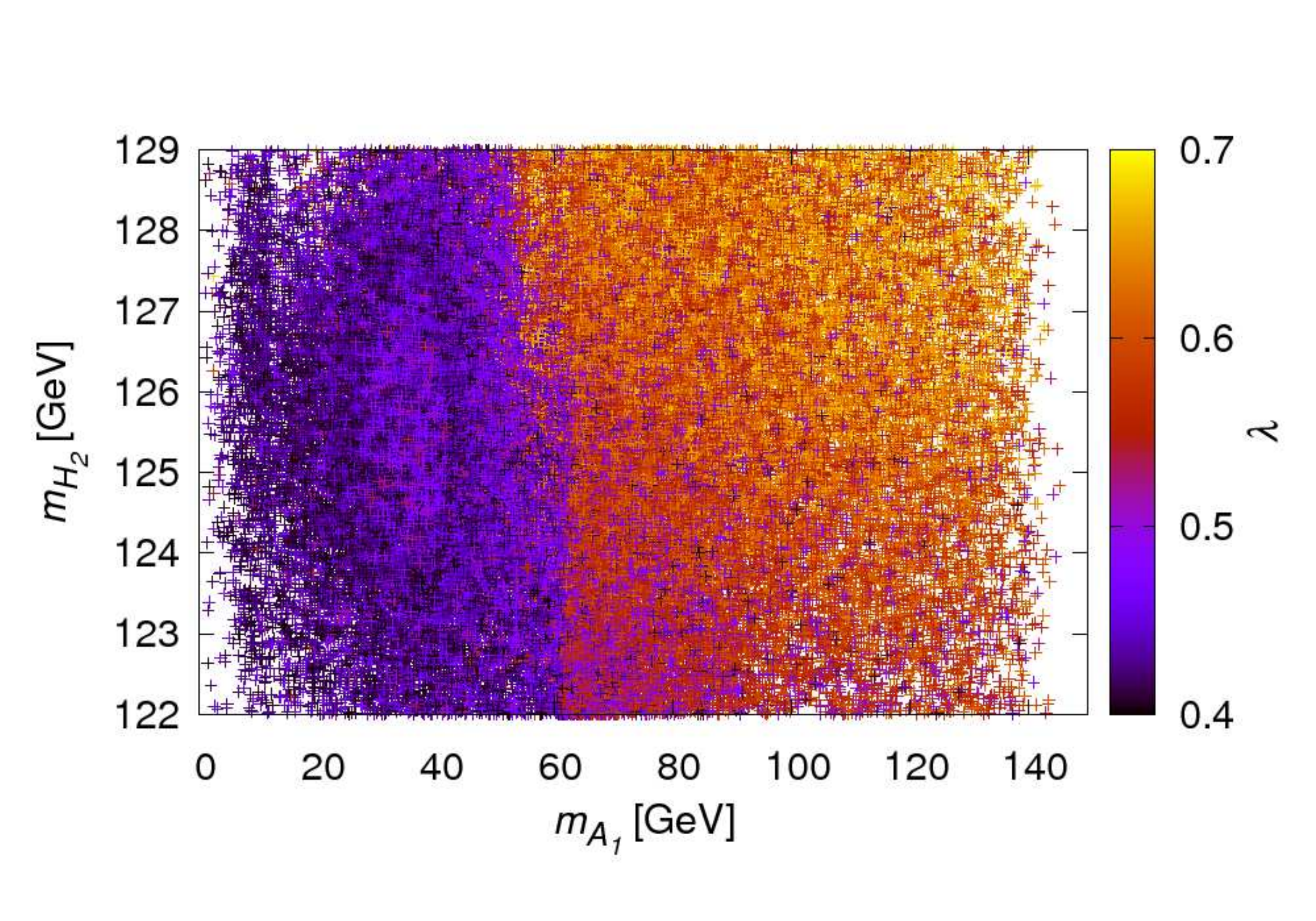}
}%
\caption[]{Mass of $H_2$ vs. that of $A_1$ for the case with $H_{\rm
    SM} = H_2$. The heat map
 shows (a) the distribution of \tanb\ and (b) the distribution of \lam.}
\label{fig:h2params1}
\end{figure}

Figs.~\ref{fig:h2params1}(a) and (b) show that $H_2$ with a mass lying in the entire allowed
range can be obtained much more easily without always requiring very low
$\tan\beta$ or very large $\lambda$. Moreover, the corresponding
parameter space points can also yield fairly 
small $A_1$ (with sizeable BR($H_2 \rightarrow A_1A_1/Z$)), without
the $H_2$ deviating too much from the LHC
Higgs boson signal rate measurements. We will, therefore, concentrate
in the remainder of this report only on the $H_2$ solution for
$H_{\rm SM}$ (i.e., $H_{\rm SM} = H_2$). 

\subsection{Production via $H_{\rm SM} \rightarrow A_1A_1/Z$}

\begin{figure}[tbp]
\centering
{%
\label{fig:-a}%
\includegraphics*[width=7.3cm]{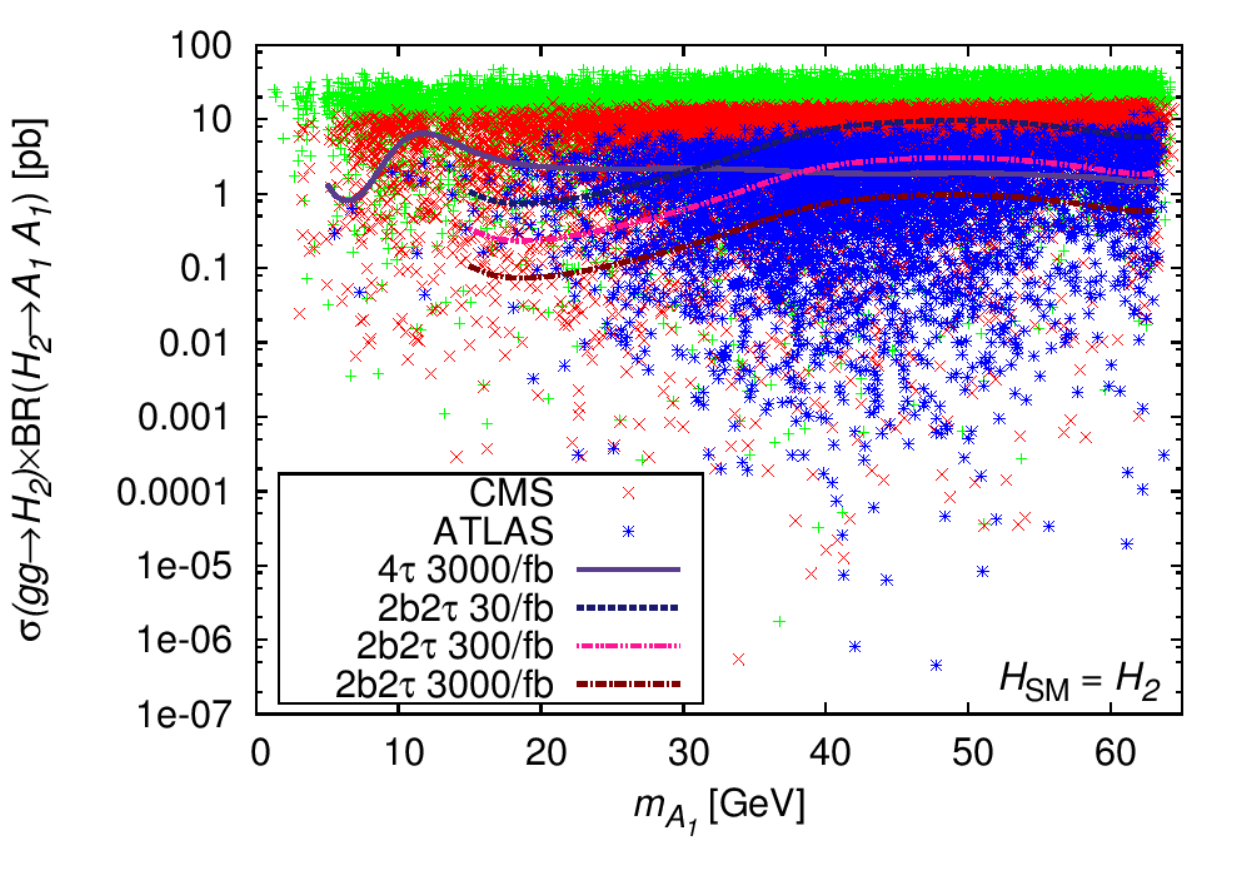}
}%
\hspace{0.5cm}%
{%
\label{fig:-b}%
\includegraphics*[width=7.3cm]{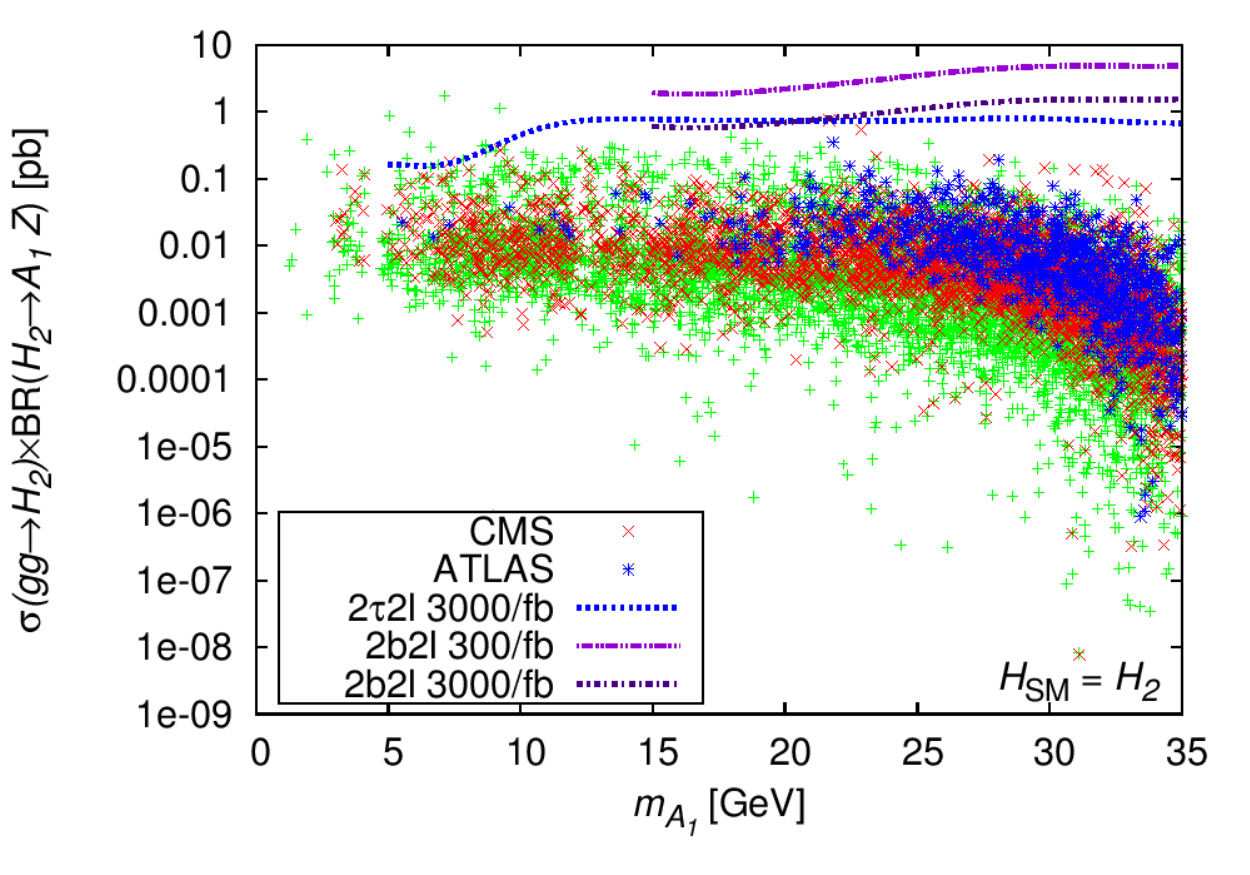}
}
\caption{Total cross sections for $H_{\rm SM} =
 H_2$ for (a) $gg \to H_{\rm SM}\to  A_1A_1$ and (b)
 $gg \to H_{\rm SM}\to A_1Z$.}
\label{fig:H2toA1}
%\end{center}
\end{figure}

In  Fig.~\ref{fig:H2toA1}(a) we show the prospects for the $H_2 \to
A_1A_1$ channel when $H_2$ is SM-like. 
We see that, compared with the
$H_{\rm SM}=H_1$ case, a much larger part of the
parameter space can be probed at the LHC, even at as
low as 30/fb of integrated luminosity. The reason is clearly that in
this case the points with $m_{A_1}<m_{H_{\rm SM}}/2$ belong to the parameter space regions where
BR($H_2 \to A_1A_1$) is indeed sufficiently enhanced
without causing the $H_{\rm SM}$ for these points to depart from a SM-like behaviour. 
This is also the reason why
a large fraction of the points with large event rates is
consistent also with the CMS and ATLAS measurements of
$\mu_{\gamma\gamma / ZZ}$. In  Fig.~\ref{fig:H2toA1}(b) we see instead that the prospects in the $H_2\to
A_1Z$ channel are poor. 

\subsection{Production via $H' \rightarrow A_1 A_1/Z$}

\begin{figure}[tbp]
\centering
{%
\label{fig:-a}%
\includegraphics*[width=7.3cm]{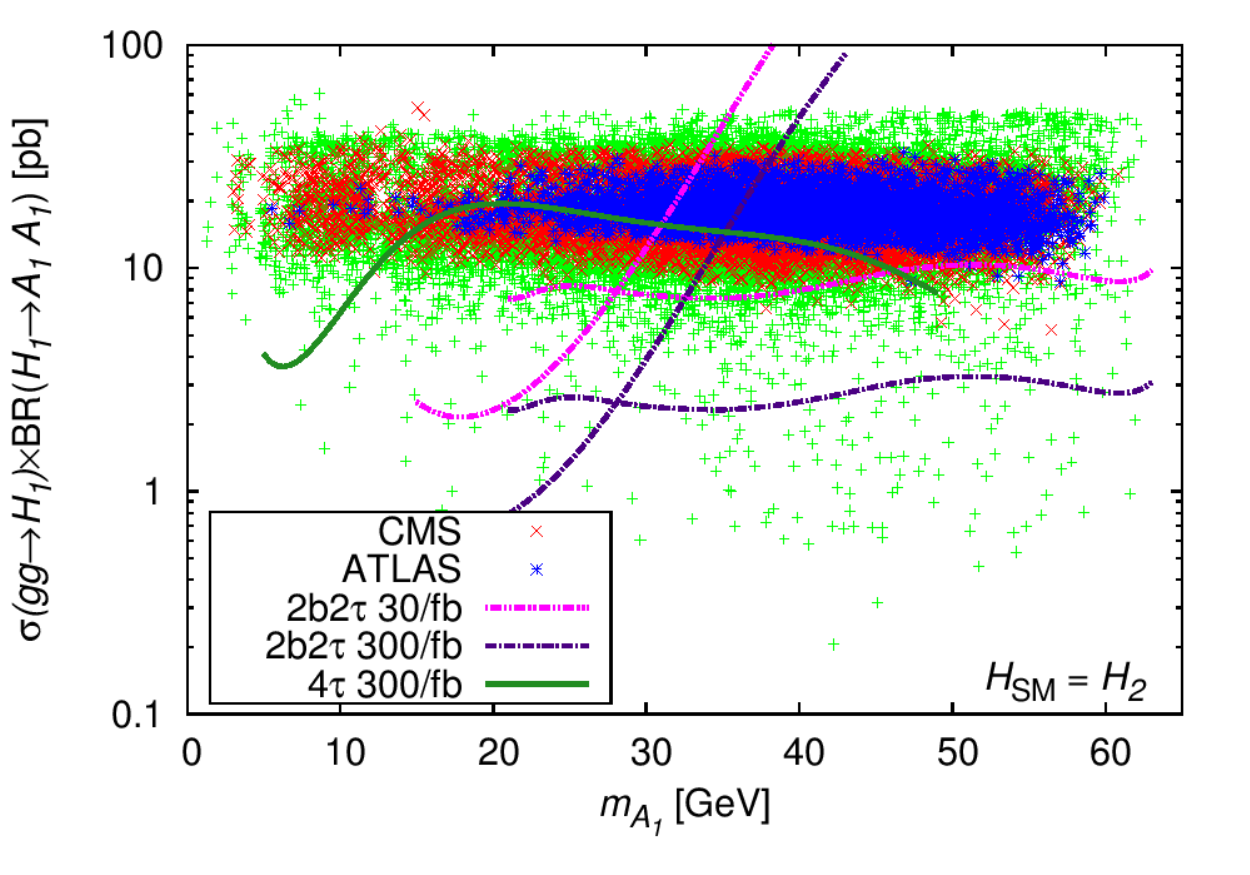}
}%
\hspace{0.5cm}%
{%
\label{fig:-b}%
\includegraphics*[width=7.3cm]{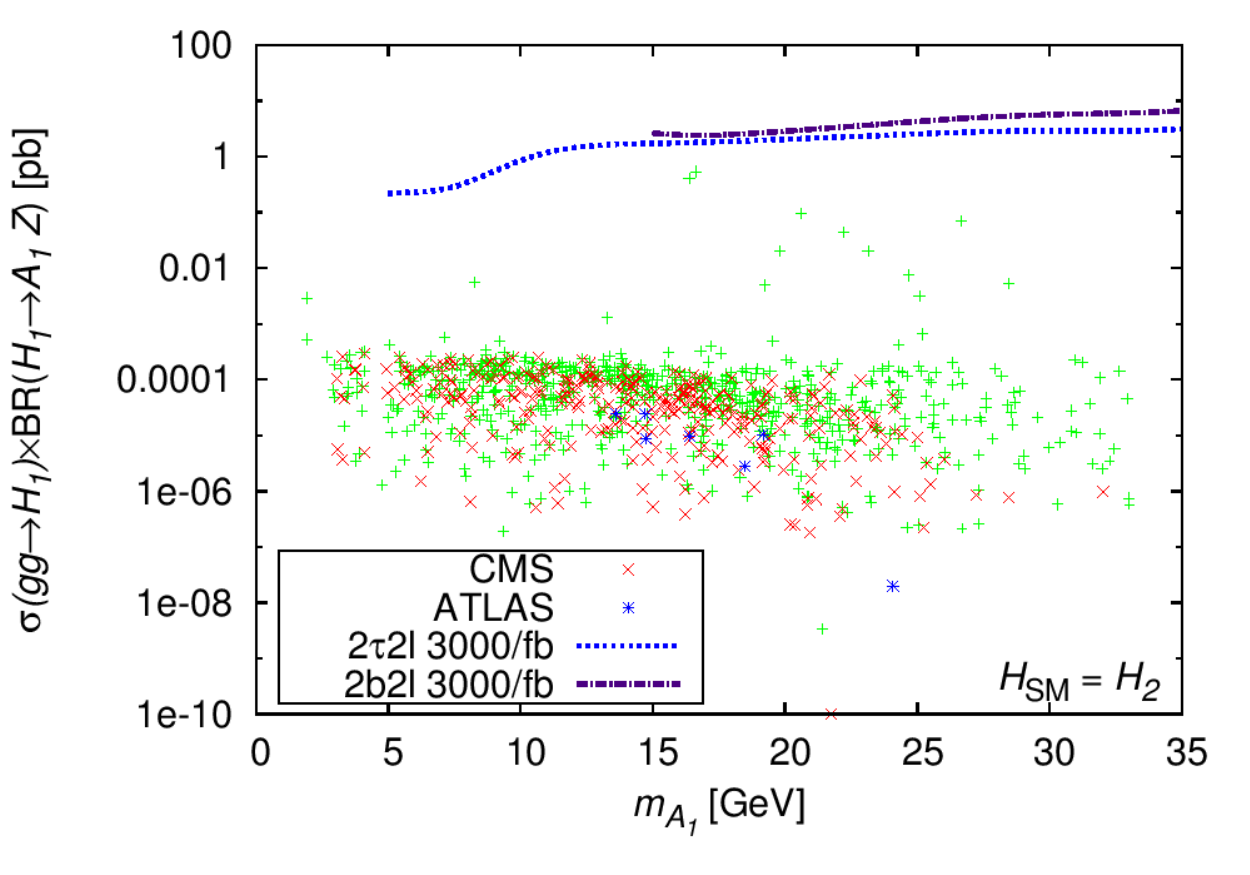}
}
\caption{Total cross sections for $H_{\rm SM} =
 H_2$ for (a) $gg \to H_1 \to  A_1A_1$ and (b)
 $gg \to H_1 \to A_1Z$.}
\label{fig:H1toA1}
%\end{center}
\end{figure}

The prospects for the discovery of a light pseudoscalar in the $H_1 \to
A_1A_1$ and $H_1 \to A_1Z$ decay channels, for a singlet-like $H_1$, are illustrated in
Figs.~\ref{fig:H1toA1}(a) and (b), respectively. One sees in  Fig.~\ref{fig:H1toA1}(a) that almost all the points
complying with the current CMS and/or ATLAS constraints on $R_X$
are potentially discoverable, even at $\mathcal{L}=30$/fb. Thus a large part
of the scanned NMSSM parameter space can be probed via this decay
channel. In particular, since such light pseudoscalars cannot easily be obtained
for the case with $H_{\rm SM}=H_1$, it should
essentially be possible to exclude or confirm $m_{A_1} \lesssim 60$\gev\ in the NMSSM at the LHC via this channel.
Note also that such an exclusion will not cover the narrow regions of parameter space where 
$m_{A_1}>m_{H_1}/2$.
Finally, In  Fig.~\ref{fig:H1toA1}(b), we see that the prospects for
the discovery of $A_1$ via the $H_1\to A_1 Z$ channel are non-existent.

\begin{figure}[tbp]
\centering
{%
\label{fig:-a}%
\includegraphics*[width=7.3cm]{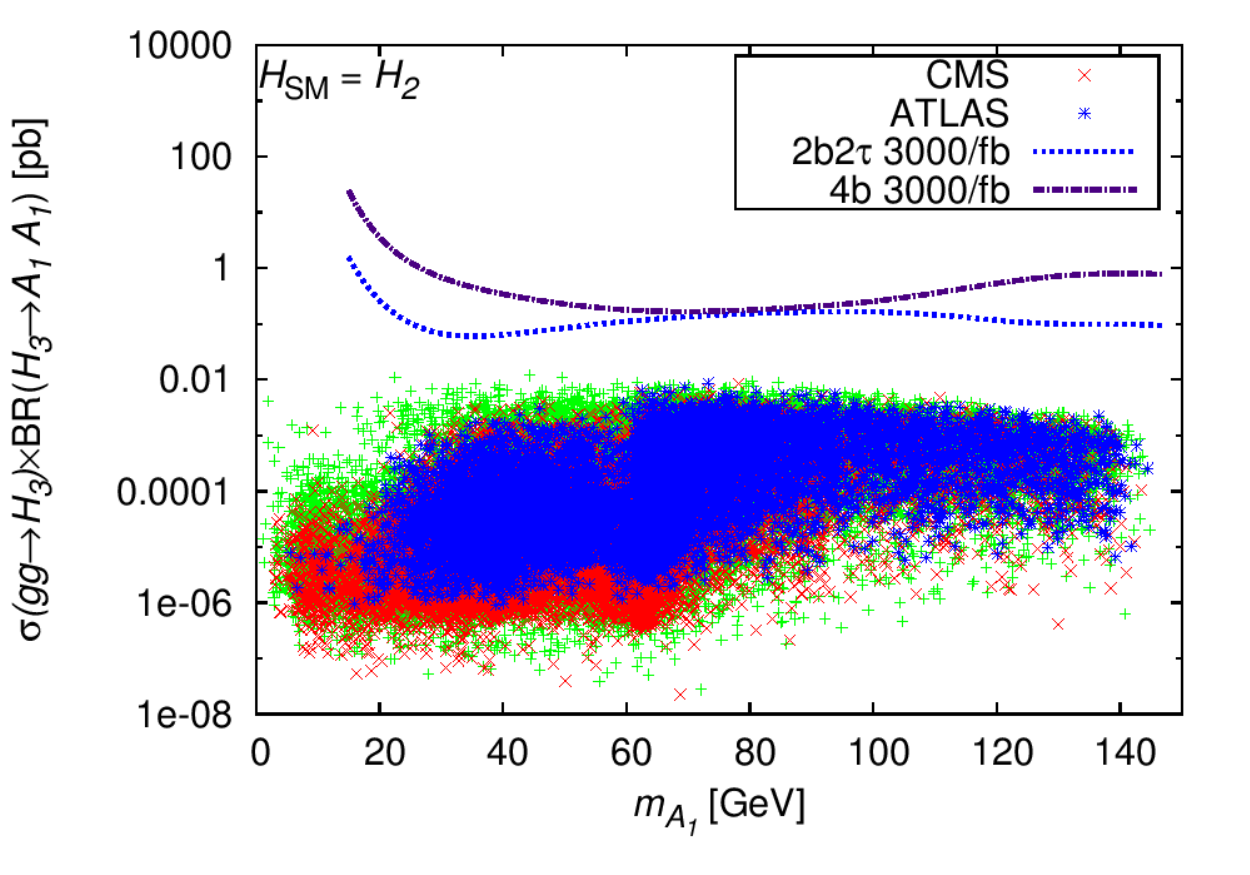}
}%
\hspace{0.5cm}%
{%
\label{fig:-b}%
\includegraphics*[width=7.3cm]{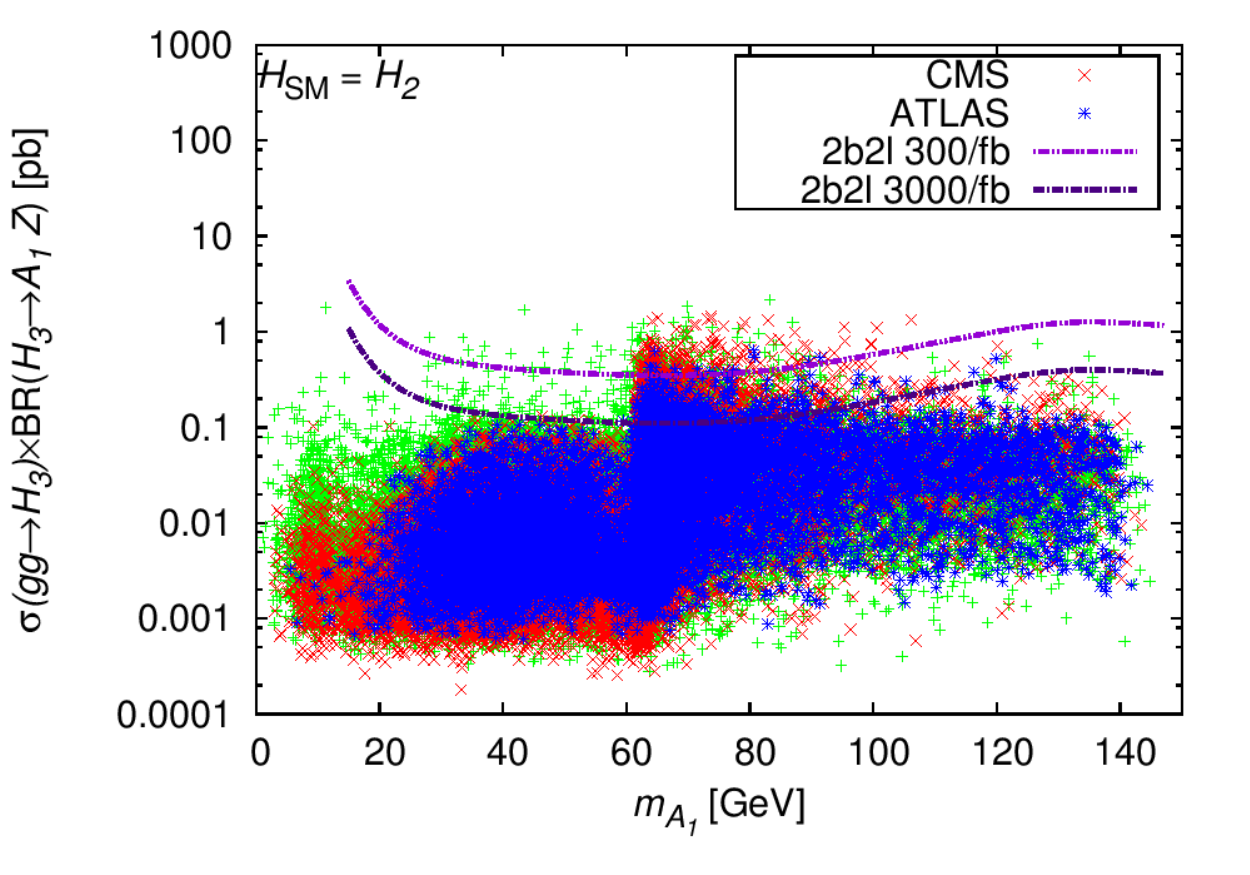}
}
\caption{Total cross sections for $H_{\rm SM} =
 H_2$ for (a) $gg \to H_3 \to  A_1A_1$ and (b)
 $gg \to H_3 \to A_1Z$.}
\label{fig:H3toA1}
%\end{center}
\end{figure}

For the decay chain starting from $H_3$, the situation is illustrated in  Fig.~\ref{fig:H3toA1}(a), where we
see that the $H_3\to A_1A_1$ channel is inaccessible  also due
to the fact that, for such high masses of $H_3$ ($\gtrsim 400$\,GeV), the
production cross section gets diminished. Moreover, other decay channels of
$H_3$ dominate. Conversely, the $H_3\to A_1Z$ channel, shown in
 Fig.~\ref{fig:H3toA1}(b), shows much more promise. This has to do with the increased sensitivity in the fat jet analysis when the involved masses are high
 as well as the relatively large $H_3A_1Z$ coupling, which is actually somewhat larger here than in the $H_{\rm SM}=H_1$ case, due to a correspondingly larger doublet
component of $A_1$. We therefore emphasise again that
this channel will be an extremely important probe for
an NMSSM $A_1$ with mass greater than $\sim 60$\gev.

\section{\label{summary} Summary}

We have found that the decays of the NMSSM CP-even scalars, including in particular
the SM-like Higgs boson, whether $H_1$ or $H_2$, carry the potential
to reveal an $A_1$ with mass $\lesssim 60$\gev\ for an integrated luminosity at the LHC
 as low as 30/fb. This is particularly true when the SM-like Higgs state is the $H_2$. 
Most notably though, when the $A_1$ is heavier than $\sim 60$\gev, while its
pair production via decays of the two lightest CP-even
Higgs bosons  becomes inaccessible, the $gg\to H_3\to A_1Z$ channel
takes over as the most promising one. This (hitherto neglected) mode  is, therefore, of great
importance and warrants dedicated probes in future analyses at the
LHC.

Such results are based upon parton-level MC event generation supplemented by parton shower and hadronisation.
Hence, final validation of our results can be done through a proper detector simulation. A key aspect of this would be the implementation of double $b$-tagging throughout.

%%%%%%%%%%%%%%%%%%%%%%%%%%%%%%%%%%
\begin{acknowledgments}
This work has been funded in part by the Welcome Programme
of the Foundation for Polish Science. S.~Moretti is financially supported in part through the NExT
Institute and further acknowledges funding from the Japan Society for the Promotion of Science (JSPS)
in the form of a Short Term Fellowship for Research in Japan
(Grant Number S14026). He is also grateful to the Theoretical Physics Group at
the Department of Physics of the University of Toyama for their  kind hospitality during the tenure of the JSPS award. 
S.~Munir is supported in part by the Swedish Research
Council under contracts 2007-4071 and 621-2011-5107. L.~Roszkowski is
also supported in part by an STFC consortium grant of Lancaster,
Manchester and Sheffield Universities. The use of the CIS computer cluster at NCBJ is
gratefully acknowledged.
% Finally, S. Moretti is grateful to the workshop organisers for a stimulating and constructive atmosphere.
\end{acknowledgments}

\end{document}

%% file: macros_gen.tex
%%%%%%%%%%%%%%%%%%%%%%% begin LR macros %%%%%%%%%%%%%%%%%%%%%%%
% 
%%%%%%%%%%%%%%%%%%

\newcommand{\newc}{\newcommand*}

% still to test
%lr \newc{\testvarmathrm}{\alpha_{\mathrm{test}}}
%lr \newc{\testvarrm}{\alpha_{\rm test}}
%lr \newc{\testvarmbox}{\alpha_{\mbox{test}}}

% For the comment, please use \begincomment & \endcomment.
\long\def\begincomment#1\endcomment{%
        \begingroup\sf\baselineskip12pt#1\endgroup}
% \long\def\begincomment#1\endcomment{}

%%%% text macros start %%%% 
\newc{\etal}{\textrm{et al.}} 
\newc{\eg}{\textrm{e.g.}} 
\newc{\ie}{\textrm{i.e.}}
\newc{\etc}{\textrm{etc}}
\newc\vs{\textrm{vs.}}
% \newc\eg{\it {e.g.}}  
% \newc\etal{\it {et al.}} 
% \newc\ie{\it i.e.}
% \newc\etc{\it {etc}}
% \newc\vs{\it {vs.}}
\newc{\cl}{\rm {CL}}

%%%% text macros  end   %%%% 

%%%% units start %%%% 
\newc{\ev}{\ensuremath{\,\mathrm{eV}}}
\newc{\kev}{\ensuremath{\,\mathrm{keV}}}
\newc{\mev}{\ensuremath{\,\mathrm{MeV}}}
%lr \newc{\gev}{\,\mathrm{GeV}}
\newc{\gev}{\ensuremath{\,\mathrm{GeV}}}
\newc{\tev}{\ensuremath{\,\mathrm{TeV}}}

%lr \newc{\GeV}{\gev}  
\newc{\MeV}{\mev} 
\newc{\TeV}{\tev}
\newc{\invpb}{\ensuremath{/\text{pb}}}
\newc{\invfb}{\ensuremath{/\text{fb}}}

\newc\nb{\ensuremath{\,\mathrm{nb}}} \newc\pb{\ensuremath{\,\mathrm{pb}}} \newc\fb{\ensuremath{\,\mathrm{fb}}}

\newc\pc{\ensuremath{\,\mathrm{pc}}}
\newc\kpc{\ensuremath{\,\mathrm{kpc}}}
\newc\mpc{\ensuremath{\,\mathrm{Mpc}}}

\newc\ps{\ensuremath{\,\mathrm{ps}}} 

% *** test this
\newc\cmeter{\ensuremath{\,\mathrm{cm}}} 
\newc\meter{\ensuremath{\,\mathrm{m}}} 
\newc\kmeter{\ensuremath{\,\mathrm{km}}}

%lr \newc\second{{\rm sec}}
%lr \newc\cmeter{{\rm cm}} 
%lr \newc\meter{{\rm m}} 
%lr \newc\kmeter{{\rm km}}
%lr \newc\second{{\rm sec}}

\newc\second{\ensuremath{\,\mathrm{s}}}
\newc\msecond{\ensuremath{\,\mathrm{ms}}}
\newc\nsecond{\ensuremath{\,\mathrm{ns}}}
\newc\psecond{\ensuremath{\,\mathrm{ps}}}

%%%% units end   %%%% 

%%%% statistics start %%%% 
%lr *** not used, changed  \chisq to \Delchisqmin *** \newc{\chisq}{\ensuremath{\chi^2 - \chi^2_{\mathrm{min}}}}
\newc{\chisqmin}{\ensuremath{\chi^2_{\mathrm{min}}}}
\newc{\Delchisq}{\ensuremath{\Delta\chi^2}}
% lr *** below changed \chsq to \chisq 
%lr \newc{\chsq}{\ensuremath{\chi^2}}
\newc{\chisq}{\ensuremath{\chi^2}}

\newc{\like}{\ensuremath{\mathcal{L}}}
%\newc{\proflike}{\ensuremath{\mathfrak L}}

%lr used in the past, not generic enough
%lr \newc{\data}\ensuremath{{d}}
%lr \newc{\nuis}\ensuremath{\psi}}
%lr \newc{\params}{\ensuremath{\theta}}
%lr \newc{\basis}{\ensuremath{m}}
%lr \newc{\derived}{\ensuremath{\xi}}
%lr \newc{\trued}{\ensuremath{\hat{\xi}}}

%%%% statistics end %%%% 

%%%% math symbols start %%%% 
\newc\lsim{\ensuremath{\mathrel{\rlap{\lower4pt\hbox{\hskip1pt$\sim$}}\raise1pt\hbox{$<$}}}}
\newc\gsim{\ensuremath{\mathrel{\rlap{\lower4pt\hbox{\hskip1pt$\sim$}}\raise1pt\hbox{$>$}}}}

\newc{\VEV}[1]{\ensuremath{\langle #1 \rangle}}

\newc{\dl}{\ensuremath{\stackrel{\leftarrow}{D}}}
\newc{\dr}{\ensuremath{\stackrel{\rightarrow}{D}}}

%%%% math symbols end   %%%% 

%%%% useful abbreviations start %%%% 

\newc{\bcenter}{\begin{center}}    \newc{\ecenter}{\end{center}}
%lr \newc{\bc}{\begin{center}}    \newc{\ec}{\end{center}}
\newc{\bfl}{\begin{flushleft}}    \newc{\efl}{\end{flushleft}}
\newc{\bfr}{\begin{flushright}}    \newc{\efr}{\end{flushright}}

\newc{\bi}{\begin{itemize}}
\newc{\ei}{\end{itemize}}
\newc{\bed}{\begin{description}}
\newc{\eed}{\end{description}}
\newc{\ben}{\begin{enumerate}}
\newc{\een}{\end{enumerate}}

\newc{\be}{\begin{equation}}
\newc{\ee}{\end{equation}}
\newc{\bea}{\begin{eqnarray}}
\newc{\eea}{\end{eqnarray}}
\newc{\ra}{\rightarrow}

%lr \newc{\beq}{\begin{equation}}
%lr \newc{\eeq}{\end{equation}}
%lr 
%lr \newc{\er}[2]{\raisebox{0.08em}{\scriptsize {$\;\begin{array}{@{}l@{}}
%lr                           \plus\makebox[0.15em][r]{#1} \\[-0.12em]
%lr                           \minus\makebox[0.15em][r]{#2}
%lr                         \end{array}$}}}
%lr \newc{\err}[2]{\raisebox{0.08em}{\scriptsize {$\;\begin{array}{@{}l@{}}
%lr                           \plus\makebox[0.55em][r]{#1} \\[-0.12em]
%lr                           \minus\makebox[0.55em][r]{#2}
%lr                         \end{array}$}}}
%lr \newc{\errr}[2]{\raisebox{0.08em}{\scriptsize {$\;\begin{array}{@{}l@{}}
%lr                           \plus\makebox[0.9em][r]{#1} \\[-0.12em]
%lr                           \minus\makebox[0.9em][r]{#2}
%lr                         \end{array}$}}}

%%%% useful abbreviations end   %%%% 

%%%% sm variable/observable start %%%% 

%lr \newc{\alphaMSbar}{\alpha_{\;{\scriptscriptstyle\protect\overline{MS}}}}
%lr examine effect of \protect\overline{MS}
\newc{\alphas}{\ensuremath{\alpha_s}}
\newc{\alphatwo}{\ensuremath{\alpha_2}}
\newc{\alphaone}{\ensuremath{\alpha_1}}
\newc{\alphai}[1]{\ensuremath{\alpha_{#1}}}
\newc{\alphaem}{\ensuremath{\alpha_{\mathrm{em}}}}

\newc{\alphaeff}{\ensuremath{\alpha_{\mathrm{eff}}}}

%\newc{\sineff}[1]{\protect{\sin2\theta_{\mathrm{eff}}^{#1}}}
\newc{\sineff}{\ensuremath{\sin \theta_{\mathrm{eff}}}}
\newc{\sinsqeff}{\ensuremath{\sin^2 \theta_{\mathrm{eff}}}}
\newc{\dalphahad}{\ensuremath{\Delta \alpha_{\mathrm{had}}}}

% Yukawa couplings
\newc{\yt}{\ensuremath{h_t}} \newc{\yb}{\ensuremath{h_b}} \newc{\ytau}{\ensuremath{h_{\tau}}}

\newc\mz{\ensuremath{M_Z}} 
\newc\mw{\ensuremath{m_W}}
\newc\mZ{\mz}        \newc\mW{\mw}

\newc\mhsm{\ensuremath{ m_{H_{\mathrm{SM}}}}}

\newc{\mtop}{\ensuremath{ m_t}}               \newc{\mtpole}{\ensuremath{ M_t}}
\newc{\mbottom}{\ensuremath{ m_b}} 
\newc{\mtau}{\ensuremath{ m_{\tau}}}
\newc{\mt}{\mtpole}
\newc{\mb}{\mbottom} 

\newc{\rtwogg}{\ensuremath{R_{h_2}(\gamma\gamma)}}
\newc{\rtwozz}{\ensuremath{R_{h_2}(ZZ)}}
\newc{\ronegg}{\ensuremath{R_{h_1}(\gamma\gamma)}}
\newc{\ronezz}{\ensuremath{R_{h_1}(ZZ)}}
\newc{\rsiggg}{\ensuremath{R_{h_\textrm{sig}}(\gamma\gamma)}}
\newc{\rsigzz}{\ensuremath{R_{h_\textrm{sig}}(ZZ)}}

\newc{\llbar}{\ensuremath{\ell\bar{\ell}}}
\newc{\tauptaum}{\ensuremath{ \tau^+\tau^-}}

\newc{\qqbar}{\ensuremath{ q\bar{q}}} \newc{\ppbar}{\ensuremath{ p\bar{p}}}
\newc{\bbbar}{\ensuremath{ b\bar{b}}} \newc{\ttbar}{\ensuremath{ t\bar{t}}}
\newc{\ffbar}{\ensuremath{ f\bar{f}}} \newc{\tautaubar}{\ensuremath{ \tau\bar{\tau}}}

\newc{\mchi}{\ensuremath{m_\neutone}}
\newc{\squark}{\ensuremath{\tilde{q}}}
\newc{\slepton}{\ensuremath{\tilde{l}}}
\newc{\gluino}{\ensuremath{\tilde{g}}} 
\newc{\mgluino}{\ensuremath{{m_{\gluino}}}}
%%%% sm variable/observable end   %%%% 

%%%% sm/susy parameters start %%%% 

\newc{\sthw}{\ensuremath{ \sin\theta_W}}              \newc{\cthw}{\ensuremath{\cos\theta_W}}
\newc{\tanthw}{\ensuremath{ \tan\theta_W}}              \newc{\cotthw}{\ensuremath{\cot\theta_W}}

\newc{\ssqthw}{\ensuremath{\sin^2 \theta_W}}

\newc{\msbar}{\ensuremath{\overline{MS}}} \newc{\drbar}{\ensuremath{\overline{DR}}}

%lr \newcommand*{\mt}\ensuremath{{m_t} %lr \newcommand*{\Mt}\ensuremath{{M_t}} %mtpole
\newc{\mtmtsmmsbar}{\ensuremath{ m_t(m_t)^{\msbar}_{{\mathrm{SM}}}}}
\newc{\mtmtsmdrbar}{\ensuremath{ m_t(m_t)^{\drbar}_{{\mathrm{SM}}}}}
\newc{\mtmtmssmdrbar}{\ensuremath{ m_t(m_t)^{\drbar}_{{\mathrm{SUSY}}}}}

%lr \newcommand*{\mb}\ensuremath{{m_b}}
\newc{\mbmbmsbar}{\ensuremath{ m_b(m_b)^{\msbar} }}

\newc{\mbmbsmmsbar}{\ensuremath{ m_b(m_b)^{\msbar}_{{\mathrm{SM}}}}}
\newc{\mbmzsmmsbar}{\ensuremath{ m_b(\mz)^{\msbar}_{{\mathrm{SM}}}}}
\newc{\mbmzsmdrbar}{\ensuremath{ m_b(\mz)^{\drbar}_{{\mathrm{SM}}}}}
\newc{\mbmzmssmdrbar}{\ensuremath{ m_b(\mz)^{\drbar}_{{\mathrm{SUSY}}}}}

\newc{\mtaumzsmmsbar}{\ensuremath{ m_{\tau}(\mz)^{\msbar}_{{\mathrm{SM}}}}}
\newc{\mtaumzsmdrbar}{\ensuremath{ m_{\tau}(\mz)^{\drbar}_{{\mathrm{SM}}}}}
\newc{\mtaumzmssmdrbar}{\ensuremath{ m_{\tau}(\mz)^{\drbar}_{{\mathrm{SUSY}}}}}

\newc{\alphasmzms}{\ensuremath{\alpha_s(M_Z)^{\overline{MS}}}}
\newc{\alphaimzms}[1]{\ensuremath{\alpha_{#1}(M_Z)^{\overline{MS}}}}

\newc{\alphaemmz}{\ensuremath{\alpha_{\mathrm{em}}(M_Z)^{\overline{MS}}}}

%%%% sm/susy parameters end   %%%% 

%%%% unified susy (cmssm, nuhm, cnmssm,...) parameters start %%%% 
\newc{\mzero}{\ensuremath{{m_0}}}
\newc{\mhalf}{\ensuremath{ m_{1/2}}}
\newc{\tanb}{\ensuremath{\tan\beta}}
\newc{\azero}{\ensuremath{ A_0}}
\newc{\signmu}{\ensuremath{\rm{sgn}\,\mu}}
\newc{\atau}{\ensuremath{{A_{\tau}}}}

%%% **** NEW *** LR: nmssm specific macros start %%%%%%
\newc{\mueff}{\ensuremath{\mu_{\rm{eff}}}}

\newc{\lam}{\ensuremath{{\lambda}}}
\newc{\kap}{\ensuremath{{\kappa}}}

\newc{\alam}{\ensuremath{{A_{\lambda}}}}
\newc{\akap}{\ensuremath{{A_{\kappa}}}}

 \newc{\hs}{\ensuremath{ H_s}}      
\newc{\mhs}{\ensuremath{ m_{H_s}}} 

%%% LR: nmssm specific macros end %%%%%%%%

\newc{\mgut}{\ensuremath{ M_{\rm GUT}}}
\newc{\mplanck}{\ensuremath{ M_{\rm P}}}      \newc{\mpl}{\ensuremath{ M_{\rm Pl}}}
\newc{\msusy}{\ensuremath{ M_{\rm SUSY}}}      \newc{\ms}{\ensuremath{ M_{\rm S}}}

%%%%%%%% nuhm %%%%%%%
 \newc{\hu}{\ensuremath{ H_u}}       \newc{\hd}{\ensuremath{ H_d}}
 \newc{\mhu}{\ensuremath{ m_{H_u}}}       \newc{\mhd}{\ensuremath{ m_{H_d}}}
 \newc{\mhuew}{\ensuremath{ m^{\ast}_{H_u}}}       \newc{\mhdew}{\ensuremath{ m^{\ast}_{H_d}}}
 \newc{\mhuewsq}{\ensuremath{ m^{\ast\, 2}_{H_u}}}       \newc{\mhdewsq}{\ensuremath{ m^{\ast\, 2}_{H_d}}}
 \newc{\mhl}{\ensuremath{m_\hl}} 
 \newc{\mhone}{\ensuremath{m_{h_1}}} 
 \newc{\mhtwo}{\ensuremath{m_{h_2}}} 
%lr \newc{\mhuast}{m^{\ast}_{H_u}}       \newc{\mhdast}{m^{\ast}_{H_d}}
 \newc{\mglu}{\ensuremath{m_{\tilde g}}} 
 \newc{\mul}{\ensuremath{m_{\tilde{u}_L}}} 
 \newc{\mtone}{\ensuremath{m_{\tilde{t}_1}}} 
 \newc{\ma}{\ensuremath{m_A}} 
 \newc{\maone}{\ensuremath{m_{a_1}}} 
 \newc{\matwo}{\ensuremath{m_{a_2}}}
 \newc{\hone}{\ensuremath{h_1}}
 \newc{\htwo}{\ensuremath{h_2}}
 \newc{\aone}{\ensuremath{a_1}}
 \newc{\atwo}{\ensuremath{a_2}}

%%%% unified susy (cmssm, nuhm, cnmssm,...) parameters  end   %%%% 

%%%%%% dark matter cosmology variables start %%%%%%%
\newc{\sigsip}{\ensuremath{\sigma^{\rm SI}_{p}}}	\newc{\sigsin}{\ensuremath{\sigma^{\rm SI}_{n}}}
\newc{\sigsdp}{\ensuremath{\sigma^{\rm SD}_{p}}}	\newc{\sigsdn}{\ensuremath{\sigma^{\rm SD}_{n}}}
\newc{\sigsi}{\ensuremath{\sigma^{\rm SI}}}	\newc{\sigsd}{\ensuremath{\sigma^{\rm SD}}}

\newc{\abund}{\ensuremath{ \Omega h^2}}
\newc{\omegadm}{\ensuremath{ \Omega_{{\rm DM}}}}     \newc{\abunddm}{\ensuremath{ \Omega_{{\rm DM}} h^2}} 
\newc{\omegam}{\ensuremath{ \Omega_{{\rm m}}}}       \newc{\abundm}{\ensuremath{ \Omega_{{\rm m}} h^2}}
\newc{\omegab}{\ensuremath{ \Omega_{{\rm b}}}}	\newc{\abundb}{\ensuremath{ \Omega_{{\rm b}} h^2}}
% \newc{\omegaB}{\ensuremath{ \Omega_{{\rm B}}}}	\newc{\abundB}{\ensuremath{ \Omega_{{\rm B}} h^2}}
\newc{\omegatot}{\ensuremath{ \Omega_{{\rm TOT}}}}
\newc{\omegacdm}{\ensuremath{ \Omega_{{\rm CDM}}}}   \newc{\abundcdm}{\ensuremath{ \Omega_{{\rm CDM}} h^2}}
\newc{\omegalambda}{\ensuremath{ \Omega_{\Lambda}}} \newc{\abundlambda}{\ensuremath{ \Omega_{\Lambda} h^2}}
\newc{\omegarad}{\ensuremath{ \Omega_{{\rm rad}}}}  \newc{\abundrad}{\ensuremath{ \Omega_{{\rm rad}} h^2}}

\newc{\rhocrit}{\ensuremath{ \rho_{\rm crit}}}
\newc{\rhochi}{\ensuremath{ \rho_{\chi}}}

\newc{\abunchi}{\ensuremath{\Omega_\chi h^2}}
\newc{\abundlsp}{\ensuremath{\Omega_{\rm LSP}h^2}}

\newcommand*{\abundchi}{\ensuremath{\Omega_\chi h^2}}% For multiple citations with one key
%%%%%% dark matter cosmology variables end %%%%%%%

%%%%%% flavor variables start %%%%%%%

% anomalous magnetic moment of the muon
\newc{\amu}{\ensuremath{ a_{\mu}}}        \newc{\amususy}{\ensuremath{ a_{\mu}^{\mathrm{SUSY}}}}
\newc{\amuexpt}{\ensuremath{ a_{\mu}^{\mathrm{expt}}}}        \newc{\amusm}{\ensuremath{ a_{\mu}^{\mathrm{SM}}}}
%lr \newc{\dasusy}{\delta a_{\mu}^{\mathrm{SUSY}}}
\newc\deltaamu{\ensuremath{\Delta a_{\mu}}} \newc{\deltaamususy}{\ensuremath{\delta a_{\mu}^{\mathrm{SUSY}}}}
\newc\gmtwo{\ensuremath{ (g-2)_{\mu}}} 
\newc{\deltagmtwomususy}{\ensuremath{\delta\left(g-2\right)_{\mu}^{\mathrm{SUSY}}}}
\newc{\deltagmtwomu}{\ensuremath{\delta\left(g-2\right)_{\mu}}}

% \newc\br{\mbox{BR}}

\newc\BR{\ensuremath{\rm BR}}

\newc\bsgamma{\ensuremath{ b\rightarrow s \gamma }}
\newc\bxsgamma{\ensuremath{\overline{B}\rightarrow X_{s}\gamma}}

\newc\brbsgamma{\ensuremath{\BR\left(\bsgamma\right)}}
\newc\brbxsgamma{\ensuremath{\BR\left(\bxsgamma\right)}}

\newc\bsmumu{\ensuremath{B_s\to\mu^+\mu^-}}
\newc\brbsmumu{\ensuremath{\BR\left(B_s\to\mu^+\mu^-\right)}}

\newc\bdmmumu{\ensuremath{\overline{B}_d\to\mu^+\mu^-}}

\newc\bbbarmix{\ensuremath{\overline{B}_s\mbox{-}B_s}}      % B_s mixing
% \newc\bbbarmix{\bar{B}_s-B_s}      % B_s mixing
\newc\delmbs{\ensuremath{\Delta M_{B_s}}}

\newc{\butaunu}{\ensuremath{B_u \rightarrow \tau \nu}}
\newc{\brbutaunu}{\ensuremath{\BR\left(B_u \rightarrow \tau \nu\right)}}

%%%%%% flavor variables end %%%%%%%

% For referencing tables, figures, equations, sections
% Redefine the \cite command to include a ~
%lr *** \let\oldcite\cite
%lr *** \renewcommand*{\cite}{~\oldcite}
\newcommand*{\reftable}[1]{Table~\ref{#1}}         \newcommand*{\reftables}[2]{Tables~\ref{#1} and \ref{#2} }
\newcommand*{\reffigure}[1]{Figure~\ref{#1}}       
\newcommand*{\reffig}[1]{Fig.~\ref{#1}}
\newcommand*{\reffigs}[2]{Figs.~\ref{#1} and \ref{#2} } 
\newcommand*{\refequation}[1]{Equation~\ref{#1}}        \newcommand*{\refeq}[1]{Eq.~\ref{#1}}
\newcommand*{\refsection}[1]{Section~\ref{#1}}     \newcommand*{\refsec}[1]{Sec.~\ref{#1}}
\newcommand*{\refref}[1]{Ref.\cite{#1}}            \newcommand*{\refcite}[1]{\cite{#1}}

% Particles
\newcommand*{\neutone}{\ensuremath{\chi}}
\newcommand*{\neuttwo}{\ensuremath{{\chi}^0_2}}
\newcommand*{\neutthree}{\ensuremath{{\chi}^0_3}}
\newcommand*{\neutfour}{\ensuremath{{\chi}^0_4}}

\newcommand*{\charone}{\ensuremath{{\chi}^{\pm}_1}}
\newcommand*{\chartwo}{\ensuremath{{\chi}^{\pm}_2}}

\newcommand*{\ETslash}{\ensuremath{/ \hspace{-.7em} E_T}}
\newcommand*{\pTslash}{\ensuremath{/ \hspace{-.7em} p_T}}
\newcommand*{\hatmh}{\ensuremath{\hat{m}_h}}
\newcommand*{\mhhat}{\ensuremath{\hat{m}_h}}
\newcommand*{\stau}{\ensuremath{\tilde{\tau}}}

% CMS cuts
\newcommand*{\alphaT}{\ensuremath{\alpha_T}}
\newcommand*{\alphaTexp}{\ensuremath{\cms\ \alphaT\ 1.1\invfb} analysis}
\newcommand*{\alphaTlim}{\ensuremath{\cms\ \alphaT\ 1.1\invfb} limit}
\newcommand*{\alphaTonefb}{\ensuremath{\cms\ \alphaT\ 1.1\invfb} }
\newcommand*{\razor}{\textrm{razor}}
\newcommand*{\razorfourfb}{\ensuremath{\cms\ \razor\ 4.4\invfb} }
\newcommand*{\razorexp}{\ensuremath{\cms\ \razor\ 4.4\invfb} analysis}
\newcommand*{\razorlim}{\ensuremath{\cms\ \razor\ 4.4\invfb} limit}
\newcommand*{\Ht}{\ensuremath{H_T}}
\newcommand*{\et}{\ensuremath{E_T}}
\newcommand*{\pt}{\ensuremath{p_T}}

% Names that require e.g. small scaps
% BayesFIT name
\newcommand*{\bayesfits}{\textrm{The BayesFits Group}}
%lr *** \newcommand*{\bayesfits}{\textit{Bayes}\textsc{Fits}}
\newcommand*{\softsusy}{SOFTSUSY}
\newcommand*{\feynhiggs}{FeynHiggs}
\newcommand*{\micromegas}{MicrOMEGAs}
\newcommand*{\multinest}{MultiNest}
\newcommand*{\superbayes}{\text{SuperBayeS}}
\newcommand*{\herwig}{\text{Herwig++}}
\newcommand*{\susyhit}{\text{SUSY-HIT}}
\newcommand*{\atlas}{\text{ATLAS}}
\newcommand*{\cms}{\text{CMS}}
\newcommand*{\cmssm}{\text{CMSSM}}
\newcommand*{\wmap}{\text{WMAP}}
\newcommand*{\pythia}{\text{PYTHIA}}
\newcommand*{\xenon}{\text{XENON100}}
\newcommand*{\superiso}{\text{SuperIso}}
\newcommand*{\superisorelic}{\text{SuperIso Relic}}
\newcommand*{\stauc}{\text{\stau-coannihilation}}
%%%
\newcommand*{\nmssmtools}{\text{NMSSMTools}}

% For referencing tables, figures, equations, sections
% Redefine the \cite command to include a ~
\let\oldcite\cite
\renewcommand*{\cite}{~\oldcite}
\newcommand*{\datalr}{d}

\newcommand*{\hl}{\ensuremath{h}}

\newcommand*{\hh}{\ensuremath{H}}
\newcommand*{\mhh}{\ensuremath{m_\hh}}

\newcommand*{\ha}{\ensuremath{A}}
\newcommand*{\mha}{\ensuremath{m_\ha}}
\newcommand*{\mh}{\ensuremath{m_h}}

\newcommand*{\hpm}{\ensuremath{H^{\pm}}}
\newcommand*{\mhpm}{\ensuremath{m_\hpm}}
%%%%%%%%%%%%%%%%%%%%%%%%%%%%%%%%%%%%%%%%%%%%%%%%%%%%%%%%%%%%%%%%%
% End of Definitions and commands
%%%%%%%%%%%%%%%%%%%%%%%%%%%%%%%%%%%%%%%%%%%%%%%%%%%%%%%%%%%%%%%%%

%% file: HPNP2015_Moretti.bbl
\bigskip

\begin{thebibliography}{99} 

\bibitem{Ellwanger:2009dp} U.~Ellwanger, C.~Hugonie and A.~M.~Teixeira, 
%\emph{The Next-to-Minimal Supersymmetric Standard Model},
  {Phys.~Rept.} {\bf 496}, 1 (2010). %  [{\tt arXiv:0910.1785}].

%\cite{Bomark:2014gya}
\bibitem{Bomark:2014gya} 
  N.~E.~Bomark, S.~Moretti, S.~Munir and L.~Roszkowski,
  %``A light NMSSM pseudoscalar Higgs boson at the LHC redux,''
  JHEP {\bf 1502}, 044 (2015).
%  [arXiv:1409.8393 [hep-ph]].
  %%CITATION = ARXIV:1409.8393;%%

%SM%\cite{Bomark:2014qua}
%SM\bibitem{Bomark:2014qua} 
%SM  N.~E.~Bomark, S.~Moretti, S.~Munir and L.~Roszkowski,
%SM%``Revisiting a light NMSSM pseudoscalar at the LHC,''
%SM  arXiv:1412.5815 [hep-ph].
%SM  %%CITATION = ARXIV:1412.5815;%%

%\cite{Almarashi:2012ri}
\bibitem{Almarashi:2012ri} 
  M.~M.~Almarashi and S.~Moretti,
  %``Scope of Higgs production in association with a bottom quark pair in probing the Higgs sector of the NMSSM at the LHC,''
  arXiv:1205.1683 [hep-ph].
  %%CITATION = ARXIV:1205.1683;%%
  %4 citations counted in INSPIRE as of 17 Feb 2015

\bibitem{preparation}
  N.~E.~Bomark, S.~Moretti and L.~Roszkowski, in preparation.

\bibitem{Kowalska:2012gs}
K.~Kowalska, S.~Munir, L.~Roszkowski, E.~M. Sessolo, S.~Trojanowski et~al.,
 % {\it {The Constrained NMSSM with a 125 GeV Higgs boson -- A global analysis}},  
{Phys. Rev. D} {\bf 87},   115010 (2013).
  %[\href{http://arxiv.org/abs/1211.1693}{{\tt arXiv:1211.1693}}].

\bibitem{NMSSMTools}
{\tt{http://www.th.u-psud.fr/NMHDECAY/nmssmtools.html}}.

\bibitem{Beringer:1900zz}
{Particle Data Group Collaboration}, %J.~Beringer et~al., 
%{\it {Review of Particle Physics  (RPP)}},  
{Phys. Rev. D} {\bf 86}, 010001 (2012).

\bibitem{Ade:2013zuv}
{Planck Collaboration}, %P.~Ade et~al., 
%{\it {Planck 2013 results. XVI. Cosmological parameters}},  
{Astron. Astrophys.} {\bf 571}, A16 (2014).
%  [\href{http://arxiv.org/abs/1303.5076}{{\tt arXiv:1303.5076}}].

\bibitem{Bechtle:2013wla}
P.~Bechtle, O.~Brein, S.~Heinemeyer, O.~Stal, T.~Stefaniak et~al., 
%{\it{$\mathsf{HiggsBounds}-4$: Improved Tests of Extended Higgs Sectors against  Exclusion Bounds from LEP, the Tevatron and the LHC}},  
{Eur. Phys. J. C} {\bf 74}, 2693 (2014).
%[\href{http://arxiv.org/abs/1311.0055}{{\tt arXiv:1311.0055}}].

\bibitem{CMS-PAS-HIG-14-009}
CMS Collaboration,
%{\it Precise determination of the mass of the higgs boson and studies of the  compatibility of its couplings with the standard model},  Tech. Rep.
  CMS-PAS-HIG-14-009. %, CERN, Geneva, Jul, 2014.

\bibitem{ATLAS-CONF-2014-009}
ATLAS Collaboration,
%{\it Updated coupling measurements of the higgs boson with the atlas detector  using up to 25\,fb$^{-1}$ of proton-proton collision data},  Tech. Rep.
  ATLAS-CONF-2014-009. %, CERN, Geneva, May, 2014.

\bibitem{Harlander:2012pb}
R.~V. Harlander, S.~Liebler and H.~Mantler, 
%{\it {SusHi: A program for the  calculation of Higgs production in gluon fusion and bottom-quark annihilation  in the Standard Model and the MSSM}},  
{Comp. Phys. Commun.}  {\bf 184}, 1605 (2013).
%[\href{http://arxiv.org/abs/1212.3249}{{\tt  arXiv:1212.3249}}].

%SM%\cite{Alwall:2011uj}
%SM\bibitem{Alwall:2011uj} 
%SM  J.~Alwall, M.~Herquet, F.~Maltoni, O.~Mattelaer and T.~Stelzer,
%SM  %``MadGraph 5 : Going Beyond,''
%SM  JHEP {\bf 1106}, 128 (2011).
%SM%  [arXiv:1106.0522 [hep-ph]].
%SM  %%CITATION = ARXIV:1106.0522;%%

%\cite{Alwall:2014hca}
\bibitem{Alwall:2014hca}
J.~Alwall, R.~Frederix, S.~Frixione, V.~Hirschi, F.~Maltoni, et~al., 
%{\it {The
%  automated computation of tree-level and next-to-leading order differential
%  cross sections, and their matching to parton shower simulations}},  {\em
  JHEP {\bf 1407}, 079 (2014). 
%[arXiv:1405.0301 [hep-ph]].
%%CITATION = ARXIV:1405.0301;%% 

\bibitem{Sjostrand:2007gs}
T.~Sjostrand, S.~Mrenna and P.~Z. Skands, 
%{\it {A Brief Introduction to PYTHIA 8.1}},  
{Comput. Phys. Commun.} {\bf 178}, 852.
%  [\href{http://arxiv.org/abs/0710.3820}{{\tt arXiv:0710.3820}}].

\bibitem{Cacciari:2011ma}
M.~Cacciari, G.~P. Salam and G.~Soyez, 
%{\it {FastJet User Manual}},  
{Eur. Phys. J. C} {\bf 72}, 1896 (2012).
%  [\href{http://arxiv.org/abs/1111.6097}{{\tt arXiv:1111.6097}}].

\bibitem{Butterworth:2008iy}
J.~M. Butterworth, A.~R. Davison, M.~Rubin and G.~P. Salam, 
%{\it {Jet substructure as a new Higgs search channel at the LHC}},  
{Phys. Rev. Lett.} {\bf 100}, 242001 (2008).
%  [\href{http://arxiv.org/abs/0802.2470}{{\tt arXiv:0802.2470}}].


\end{thebibliography}
